\documentclass[journal=jpcl,manuscript=article]{achemso}
\usepackage{graphicx}
\usepackage{amsmath}
\usepackage{amssymb}
\usepackage{bm}

\author{Josep Planelles}
\affiliation{Departament de Qu\'{\i}mica F\'{\i}sica i Anal\'{\i}tica,
Universitat Jaume I, E-12080, Castell\'o de la Plana, Spain}
\email{josep.planelles@uji.es}
\author{Fernando Rajadell}
\affiliation{Departament de Qu\'{\i}mica F\'{\i}sica i Anal\'{\i}tica,
Universitat Jaume I, E-12080, Castell\'o de la Plana, Spain}
\author{Juan I. Climente}
\affiliation{Departament de Qu\'{\i}mica F\'{\i}sica i Anal\'{\i}tica,
Universitat Jaume I, E-12080, Castell\'o de la Plana, Spain}

\date{\today}

\title{Electronic Origin of Linearly Polarized Emission in CdSe/CdS Dot-in-Rod Heterostructures}


\begin{document}

\newpage

\begin{abstract}
Seeded CdSe/CdS nanorods exhibit intense polarized emission along the rod main axis.
The degree of linear polarization cannot be explained by dielectric effects alone, an additional electronic
contribution is present whose nature has not been settled up to date. Using multi-band k$\cdot$p theory, we analyze 
the potential influence of several factors affecting excitonic emission 
 and show that shear strain is the main electronic mechanism promoting linear polarization.
It favors energetically light hole excitons over heavy hole ones, via deformation potential, 
and makes their radiative recombination faster via piezoelectricity.
Implications of this mechanism are that linear emission can be enhanced by growing long but thin CdS shells
around large, prolate CdSe cores, which indeed supports and rationalizes recent experimental findings.
Together with the well-known dielectric effects, these results pave the way for controlled 
degree of linear polarization in dot-in-rods through dedicated structural design.
\end{abstract}



\newpage


The optoelectronic properties of semiconductor colloidal nanocrystals can be widely tuned by
tailoring their size, shape and composition, which makes them competitive building blocks for
exisiting and emerging technologies alike.\cite{KovalenkoACS}
One of the most promising types of nanocrystals for room temperature optical applications
in the visible spectrum are heterostructured CdSe/CdS dot-in-rods (DiRs), 
where a rod-shaped CdS shell is grown around a nearly spherical CdSe core.\cite{SittNT} 
Added to a well established synthetic route,\cite{TalapinNL,CarboneNL,CoropceanuACS}
CdSe/CdS DiRs exhibit narrow emission line width,\cite{MullerPRL} high quantum yield,\cite{TalapinNL,CarboneNL,CoropceanuACS}
optical gain,\cite{SabaAM,DiStasioSMALL} tunable Stokes shifts,\cite{TalapinNL}
tunable exciton lifetime\cite{RainoACS} and large two-photon absorption cross section.\cite{AllioneACS}

As compared to other CdSe/CdS nanocrystals, like giant-shell quantum dots\cite{ChenJACS}, 
DiRs present the additional advantage of displaying linearly polarized band edge absorption
and emission.\cite{TalapinNL,CarboneNL,CoropceanuACS}
The origin of such a linear polarization is however a controversial issue.
It is widely accepted that there is a contribution arising from the dielectric
mismatch between the inorganic structure and its surface ligands.
Because of the anisotropic shape, dielectric screening of electromagnetic fields 
favors light absorption and emission along the long axis of the rod, similar to the case of
CdSe nanorods.\cite{ShabaevNL,KamalPRB}
But the dielectric effect predicts the degree of linear polarization increase monotonically 
with the aspect ratio of the nanocrystal, saturating for aspect ratios above $\sim 10$.
This is in contradiction with a number of experiments, which have reported 
non-monotonic dependence on the aspect ratio\cite{VezzoliACS}
and polarization degrees well beyond the limit of the dielectric model.\cite{DirollJPCL,SittNL,HadarJPCL}
An additional source of optical anisotropy must be present, 
presumably related to electronic degrees of freedom.
By analogy with CdSe nanorods\cite{HuSCI}, it has been postulated that the top of the valence
band in DiRs is formed by light holes (LH) instead of heavy holes (HH), as the former promote
emission along the $c-$axis of wurtzite nanocrystals, which is aligned with the rod long axis. 
Why this occurs is still unclear, because band edge exciton recombination in DiRs takes place inside the 
core, that is nearly spherical. One would then expect the ground state to be HH.
In fact, although wurtzite CdSe seeds are slightly prolate, they do not show steady state optical
anisotropy by themselves.\cite{SittNL} Encapsulation within an anisotropic CdS shell seems to be a requiste to 
that end.

Recent experiments have attempted to elucidate the role of different morphological and 
structural parameters in determining the degree of linear polarization.\cite{SittNL,HadarJPCL,DirollJPCL,VezzoliACS}
Sitt {\it et al.} showed that elongated (rod shaped) cores strengthen linear polarization as compared
to usual (nearly-spherical) ones.\cite{SittNL,HadarJPCL}
Diroll {\it et al.} suggested that thin anisotropic shells around the core further favor
linear polarization,\cite{DirollJPCL} but Vezzoli {\it et al.} pointed out this could be 
simply due to the large size of the CdSe cores in such samples, as in  their work
large cores seem to promote LH ground states and hence linearly polarized emission.\cite{VezzoliACS} 
However, other works present DiRs with equally large core sizes and yet they report HH ground states.\cite{GranadosACS}
The precise shape of the core may make the difference.\cite{VezzoliACS}
All of these are phenomenological observations inferred from systematic experiments, 
which do not answer the fundamental question of what quantum-physical mechanism makes DiRs emit 
net linearly polarized light.
Deeper theoretical understanding is needed to unravel the influence of the 
many competing factors. This is ultimately needed to design DiRs with deterministic control of 
the optical anisotropy. The goal of this work is then to clarify the origin of the electronic contribution 
to the linear polarization of CdSe/CdS DiRs, and how it relates to the structural parameters.
 


\begin{figure}[h]
\includegraphics[width=0.8\textwidth]{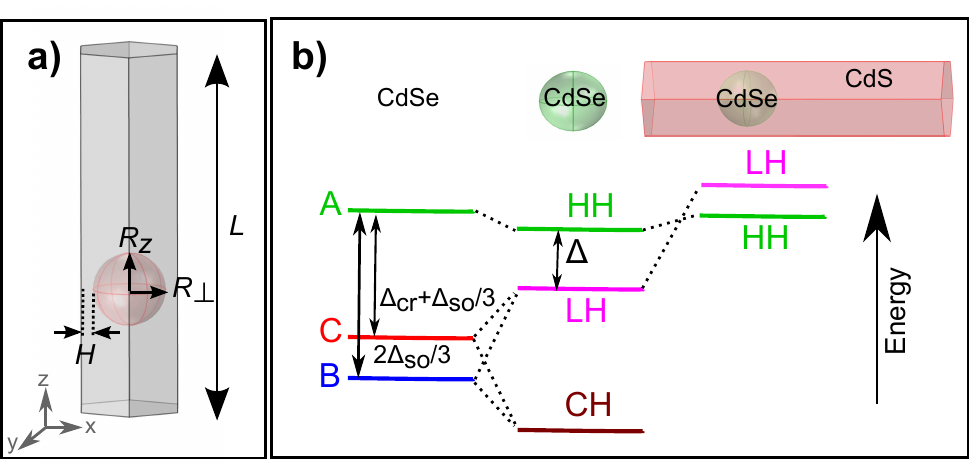}
\caption{(a) Schematic of the DiR under study. (b) Evolution of valence subbands in 
different cases. Left: uncoupled wurtzite subbands. Center: subbands coupled by confinement
and spin-orbit interaction. Right: subbands in a DiR with LH ground state. $\Delta$ is the
net splitting between HHs and LHs.}
\label{fig1}
\end{figure}

We consider DiRs like that in \ref{fig1}(a), with an ellipsoidal CdSe core of radius $R_z$
along the $z$-direction ($c$-axis) and $R_\perp$ in the $x,y$ directions. The core is at 1/3
of the CdS shell height, whose length is $L$ and has its $c$-axis also aligned along $z$. 
The shell thickness between the core and the lateral sides of the shell is denoted as $H$.
Both CdSe and CdS have wurzite crystal structure. We then calculate valence band holes using a
6-band k$\cdot$p Hamiltonian.\cite{ChuangPRB,ClimenteJAP} In this picture, the valence band splits
into $A$ subband (Bloch functions $|P_+ \uparrow\rangle,\,|P_- \downarrow\rangle$), 
$B$ subband ($|P_+ \downarrow\rangle,\,|P_- \uparrow\rangle$) and
$C$ subband ($|P_z \uparrow\rangle$ and $|P_z \downarrow\rangle$).
$C$-subband holes are responsible for the linearly polarized absorption and emission along the $z$-direction,
because when recombining radiatively with electrons (Bloch functions $|S\uparrow\rangle$ and $|S\downarrow\rangle$),
the dipole moment element $\langle S \uparrow| \mu_z | P_X \uparrow\rangle$ is only finite if $X=Z$.

In CdSe, $A$, $B$ and $C$ subbands are energetically sorted as shown on the left of \ref{fig1}(b).
The top of the valence band is formed by $A$ holes, which are split from $C$ and $B$ holes by 
the crystal field splitting $\Delta_{cr}$ and the split-off spin-orbit splitting $\Delta_{so}$.
In a nearly-spherical CdSe quantum dot, confinement and spin-orbit interaction lead to some changes, 
as shown in the center of \ref{fig1}(b). $A$-subband remains on top, its mixing with other subbands being small.
By analogy with zinc-blende structures, it is often referred to as HH subband. 
By contrast, $B$- and $C$-subbands undergo severe coupling to form LH and split-off (CH) subbands. 
Notice that LHs inherit from $C$ holes the capacity to emit linearly polarized light.
We shall label $\Delta$ the energy splitting between the top-most HH state 
(hole state with predominant HH character) and the top-most LH state (predominant LH character),
 i.e. $\Delta=E_{HH}-E_{LH}$.
 Upon growth of an anisotropic CdS shell around the dot, experiments suggest the ground state can change
from HH to LH, as shown on the right of \ref{fig1}(b). This means $\Delta$ changes from positive to negative.
 We shall try to determine which physical mechanism and structural conditions make this possible.
 To this end, we decompose $\Delta$ as:
\begin{equation}
\Delta = \Delta_{int} + \Delta_{shape} + \Delta_{strain} + \Delta_{coulomb}. 
\label{eq1}
\end{equation}
\noindent Here $\Delta_{int}$ is the intrinsic splitting at $k=0$, owing to spin-orbit interaction
and the hexagonal lattice crystal field splitting. For CdSe, $\Delta_{int}=23.4$ meV, which implies
the ground state in bulk is HH.\cite{ChuangPRB} $\Delta_{shape}$ is the shape splitting, which accounts for 
quantum confinement and spontaneous polarization.
The latter introduces a small electric field along the $c$-axis,\cite{LiPRL} whose influence we find to be
minor. By contrast, quantum confinement can be important. 
The anisotropic masses of the valence band stabilize HHs when confinement along the $c$-axis is comparable or stronger
than the transversal one, and LHs otherwise.\cite{EfrosPRB}  Because wurtzite cores tend to deviate from 
sphericity and become slightly prolate along the $c$-axis, this could favor LHs.
For cores with sufficiently large aspect ratio, it could even lead to a LH ground state.\cite{PlanellesJPCc} 
$\Delta_{strain}$ is the splitting induced by strain via deformation potential (DP) and piezoelectricity (PZ).
Strain-induced DP is felt differently by HH and LHs.\cite{ChuangPRB} 
Indeed, experiments with CdSe/CdS dot-in-plates reported unusually large and positive values of $\Delta$,
which were interpreted using simple models to be a consequence of the anisotropic pressure exerted 
by the shell upon the core.\cite{CassetteACS} Because DiRs are the opposite case to dot-in-plates,
with prolate shells instead of oblate ones, it has been speculated this could be a key mechanism 
leading to small or negative $\Delta$.\cite{VezzoliACS} We will try to verify this point.
Piezoelectricity, in turn, can stretch and separate electron and hole wave functions along the $c$-axis of CdSe/CdS
heterocrystals,\cite{ChristodoulouNC,SegarraJPCL} which may also affect the HH-LH splitting.
$\Delta_{coulomb}$ represents the influence of excitonic Coulomb attraction. Since the electron in DiRs
is partly delocalized over the shell,\cite{LuoACS,EshetNL} it may as well stretch the hole wave function
along the $c$-axis, hence stabilizing LHs. 

\begin{figure}[h]
\includegraphics[width=0.8\textwidth]{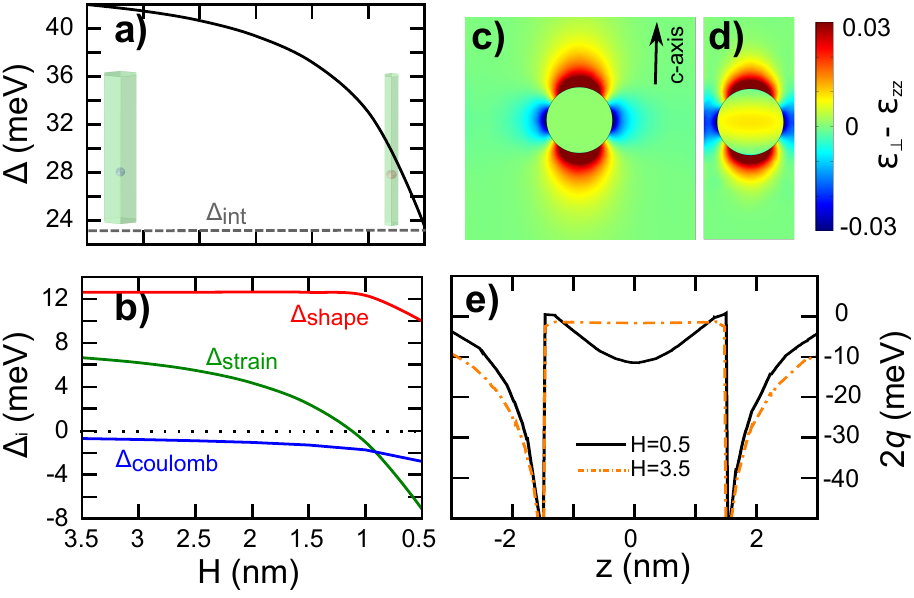}
\caption{(a) HH-LH splitting as a function of the shell thickness. Dashed lines show the intrinsic splitting $\Delta_{int}$, for comparison.
(b) Influence of individual factors on $\Delta$.  Dotted lines highlight the zero. 
(c) shear strain in and around the core of a thick-shell DiR. (d) same in a thin-shell DiR. Note the finite strain inside the core.
(e) Offset between deformation potentials seen by HH and LHs along the rod axis. 
In the thin shell case ($H=0.5$ nm), the difference inside the core becomes negative, thus favoring LHs. 
In all cases the core is spherical with $R=1.5$ nm, and rod length is $L=50$ nm.}
\label{fig2}
\end{figure}

We start by analyzing the influence of the shell dimensions on $\Delta$. We consider DiRs with a fixed spherical core of 
radius $R=1.5$ nm, shell length $L=50$ nm, and varying shell thickness $H$. 
As can be seen in \ref{fig2}(a), the net HH-LH splitting rapidly decreases as the shell becomes thinner.
This qualitatively supports the conclusion inferred from recent experiments, that thin shells enhance 
optical anisotropy.\cite{DirollJPCL} Notice, however, that the experiments included an inverse correlation 
between shell thickness and core size, which made it difficult to disentangle the influence of the two factors. 
Since we have fixed core radius, our calculations allow us to confirm the role of $H$ unambiguously.
 Notice also that in spite of the decrease, $\Delta$ remains positive even for the thinnest shells.
This suggests that thin shells alone cannot justify a LH ground state. Further factors are needed, 
which we will discuss below. Before that, it is useful analyzing why a thin shell reduces $\Delta$.
To this end, in \ref{fig2}(b) we plot the dependence of different terms of Eq.~(\ref{eq1}) on $H$.
 $\Delta_{shape}$ is roughly constant except for very thin shells. This is because the hole is mainly confined
 inside the core and hence little sensitive to changes of the shell confinement potential.
 $\Delta_{coulomb}$ is also quite unsensitive to $H$, its value being small in all cases.
 The main change is observed in $\Delta_{strain}$, which is positive for thick shells but rapidly 
decreases as $H$ gets smaller, reaching negative values for $H<1$ nm. 
This behavior is essentially due to the DP, with PZ playing a secondary role,
and can be interpreted as follows. Within the quasi-cubic approximation for wurtzite structures,\cite{ChuangPRB,XiaPRB} 
the DP term splitting HH and LH states is (see Supporting Information):
\begin{equation}
2q=2\,d/\sqrt{3}\,\left( \epsilon_{\perp} - \epsilon_{zz} \right) \approx \Delta_{strain}, 
\end{equation}
\noindent where $d$ is a cubic deformation potential parameter, $\epsilon_{\perp}=1/2\,(\epsilon_{xx}+\epsilon_{yy})$
and $\epsilon_{zz}$ strain tensor elements orthogonal and parallel to the $c$-axis.
The shear term $(\epsilon_{\perp}-\epsilon_{zz})$ gives the difference between the vertical and in-plane lattice
constant due to the strain. Since $d<0$ for both CdSe and CdS (see Supporting Information), the splitting
decreases for $(\epsilon_{\perp}-\epsilon_{zz})>0$ and increases otherwise. 
As shown in \ref{fig2}(c), when the shell is thick, shear strain inside the spherical core is zero.
Because the hole wave function is mostly confined inside the core,\cite{LuoACS} 
this means $\Delta_{strain}$ should be negligible. In \ref{fig2}(b), we observe a small but finite $\Delta_{strain}\approx 5$ meV,
which is due to the wave function leaking into the CdS shell, where shear strain is very pronounced.
The relevant change, however, takes place when the shell is made thinner.
As shown in \ref{fig2}(d), for small $H$ values, shear strain becomes positive inside the core.
This is because the CdSe core, which is compressed inside the CdS shell, finds it easier to dilate 
in the thin shell direction. Then, $|\epsilon_\perp|$ becomes smaller than in the thick shell case,
while $|\epsilon_{zz}|$ stays the same. Because both $\epsilon_\perp$ and $\epsilon_{zz}$ are negative
(compressive), $(\epsilon_{\perp}-\epsilon_{zz})>0$. In other words, strain relaxation in the direction
orthogonal to the $c$-axis, leads to negative $\Delta_{strain}$ values stabilizing LH states. 
This is further illustrated in \ref{fig2}(e), which compares $2q$ along the $c$-axis for thick ($H=3.5$ nm)
and thin ($H=0.5$ nm) shells. When using thin shells, the DP energy offset between HH and LH 
inside the core decreases by more than 10 meV.

The influence of the shell length $L$ on $\Delta$ is much weaker than that of $H$ (see Fig.~S1 in Supporting Information).
Its main effect is to stabilize HH with respect to LHs when the shell changes from rod to disk shape
(oblate instead of prolate), which actually agrees with previous results for dot-in-plates.\cite{CassetteACS} 

\begin{figure}[h]
\includegraphics[width=0.8\textwidth]{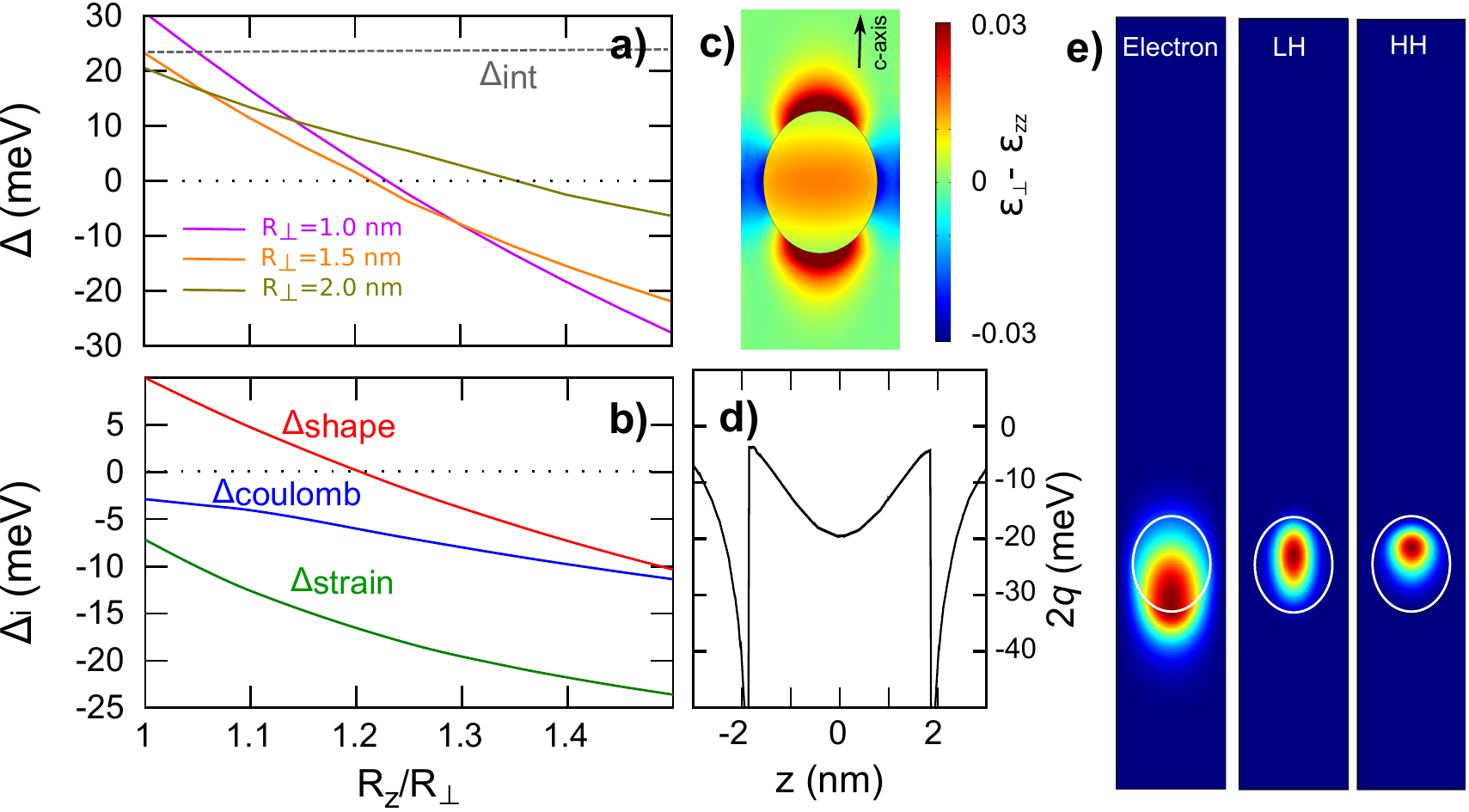}
\caption{(a) HH-LH splitting as a function of the core aspect ratio for different diameters.
(b) Influence of individual factors on $\Delta$. Dotted lines highlight the zero. 
(c) shear strain in and around the core of a prolate-core DiR. 
(d) Offset between deformation potentials seen by HH and LHs along the rod axis. 
Notice the value is the center of the core is about twice that of the spherical core in \ref{fig2}(e).
(e) Excitonic electron, HH and LH charge densities in a DiR.
In all cases the shell has thickness $H=0.5$ nm and length $L=50$ nm. 
For panels (b)-(e), the core has $R_\perp=1.5$ nm. For panels (c)-(e), $R_z/R_\perp=1.25$.} 
\label{fig3}
\end{figure}

 As mentioned above, in spite of the negative $\Delta_{strain}$ value, the ground state remains HH 
even for the thinnest shells studied in \ref{fig2}. 
 In order to obtain a LH ground state, one needs to consider the additional effect of the core size and shape, 
as we show next.
We take a fixed, thin shell and vary the core radii $R_z$ and $R_\perp$.
The resulting $\Delta$ is shown in \ref{fig3}(a). 
For spherical cores ($R_z/R_\perp=1$), $\Delta>0$ regardless of the core size, but it rapidly decreases
with increasing aspect ratio of the core.
For $R_z/R_\perp$ between $1.2$ and $1.35$, depending on the core size, $\Delta$ becomes negative.
This confirms that LH ground states are feasible, but they require slightly prolate cores rather
than anisotropic shells. 
 
The fast decrease of $\Delta$ with the core aspect ratio can be understood from \ref{fig3}(b).
As the core becomes more prolate, $\Delta_{shape}$ decreases because of the anisotropic masses of HH and LH.\cite{EfrosPRB}
Yet, this term by itself would not suffice to compensate for $\Delta_{int}$ and obtain a LH ground state.
There is an additional negative contribution of $\Delta_{strain}$, which is again due to the 
different DP seen by HHs and LHs. If we compare a spherical core ($R_z/R_\perp=1$, \ref{fig2}(d))
with a slightly prolate one ($R_z/R_\perp=1.25$, \ref{fig3}(c)), it can be seen that the shear strain
inside the core is doubled. The same can be observed in the DP offset profile along the rod axis,
\ref{fig3}(d), which is twice deeper inside the core as compared to its spherical counterpart 
(black line in \ref{fig2}(e)). It follows that a small deviation from sphericity has an important
effect on $\Delta_{strain}$. 
\ref{fig3}(b) also shows that Coulomb interaction provides a negative contribution
to $\Delta$, which is enhanced with increasing core aspect ratio. The reason can be inferred from 
the charge densities plotted in \ref{fig3}(e). The exciton electron and hole charge densities
are slightly pushed towards opposite sides of the core along the $c$-axis, owing to PZ.\cite{SegarraJPCL} 
Because LHs have lighter mass in this direction, they are more elongated and their overlap with
the electron is larger than that of HHs. Consequently, electron-LH Coulomb attraction is stronger
than electron-HH one. With increasing core size and ellipticity, this effect becomes more important 
and helps stabilize LHs. 

Since wurtzite cores tend to grow prolate, all the above effects are expected to contribute in
the formation of LH ground states. The critical aspect ratio at which the ground state crossover takes
place depends on the core diameter and shell thickness, but we stress small values suffice. In fact,
we have checked that using quasi-cubic masses instead of the bulk wurtzite values 
employed in \ref{fig3}(a), the critical aspect ratio becomes even smaller, $R_z/R_\perp=1.15$ or less 
(see Fig.~S2 in Supporting Information). Likewise, using different sets of DP parameters within 
experimental uncertainty, the critical aspect ratio remains similar or slightly smaller (Fig.~S3 in Supporting Information).
 Please note that the CdS shell plays an important role on the HH-LH reversal. 
The critical aspect ratio for a CdSe quantum dot is systematically greater than that of a CdSe seed forming a DiR
(Fig.~S4 in Supporting Information).
 The fact that large cores are more prone to grow prolate could partly explains 
the experimental observation that large core DiRs show higher degrees of linear polarization than small ones.\cite{VezzoliACS}
 On a different note, the seeming contradiction between Ref.~\cite{VezzoliACS} and Ref.~\cite{GranadosACS}
experiments, which reported LH and HH ground states for DiRs with similar core radius ($R\approx 3.2$ nm)
and shell thickness ($H\approx0.5$ nm), respectively, is possibly due to the pronounced prolate shape of the former
 (see large dots in Fig.~1 of their Supporting Information), as opposed to the quasi-spherical shape
claimed by the latter. This would be consistent with our prediction that a thin shell does not suffice to
induce a LH ground state, core ellipticity being a requirement.\\

\begin{figure}[h]
\includegraphics[width=0.8\textwidth]{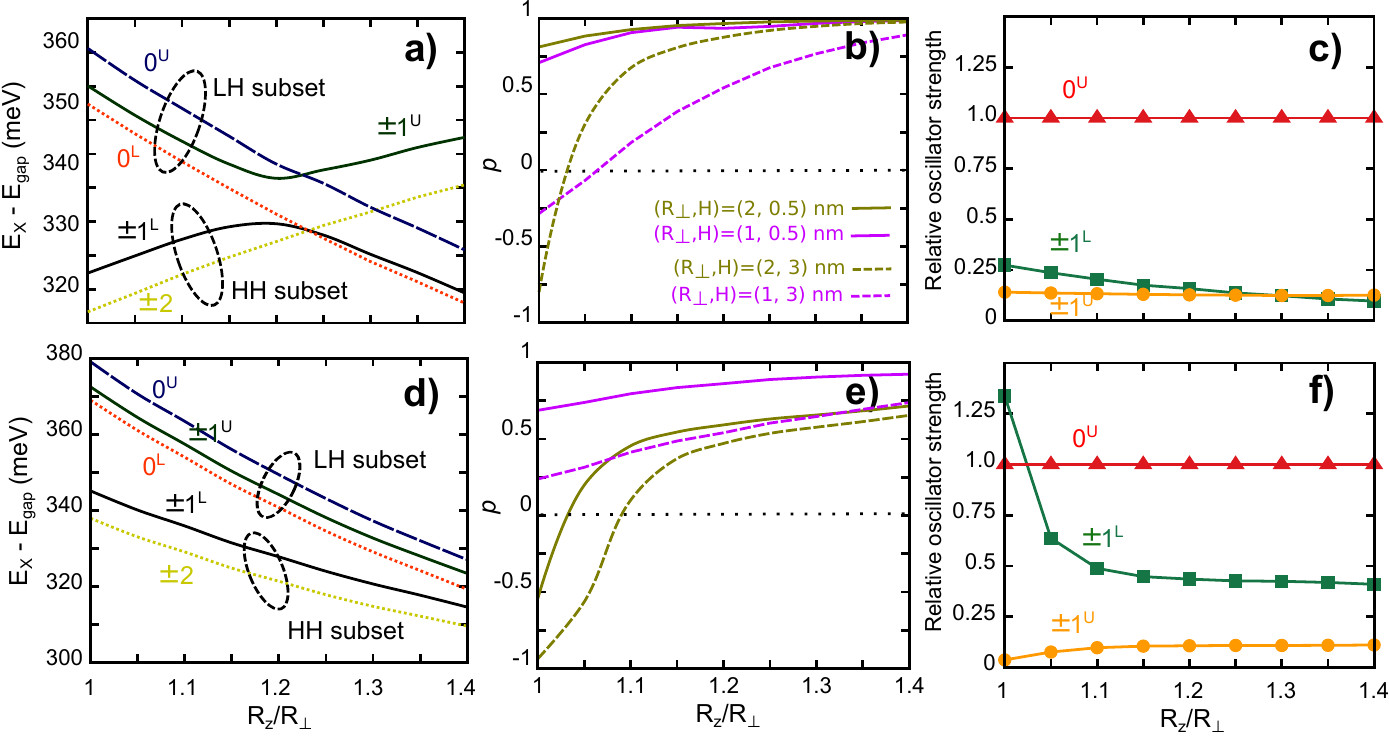}
\caption{(a) Exciton fine structure dependence on the core aspect ratio for $R_\perp=1$ nm and $H=0.5$ nm.
 HH and LH subsets cross at an aspect ratio governed by $\Delta$. 
$0^U$ is the only state emitting linearly polarized light. $\pm 1^{U/L}$ states emit planar polarized light. 
Other states (dotted lines) are dark. 
(b) Degree of linear polarization as a function of the DiR core aspect ratio, for different core 
sizes and shell thicknesses, at room temperature.
Highest $p$ values correspond to large, prolate cores with thin shells.
 (c) Oscillator strength of $\pm 1^{U/L}$ states relative to that of $0^U$ in 
DiRs with $R_\perp=2$ nm and $H=0.5$ nm. They are much weaker.
(d-f) is the same as (a-c) but neglecting strain. 
In all the plots the shell has aspect ratio of 10, so that dielectric effects are constant.
}
\label{fig4}
\end{figure}

 To see how the HH-LH energy splitting actually affects the degree of linear polarization, 
 we need to include the exciton fine structure induced by electron-hole exchange interaction. 
 We calculate this term with a wurtzite Hamiltonian (see Supporting Information), but
the results are in close agreement with the quasi-cubic approximation used in the literature.\cite{EfrosPRB2,RainoACS2}
As an example, \ref{fig4}(a) shows the exciton spectrum corresponding to a DiR with small core ($R_\perp=1$ nm) and thin shell.
In the figure, we use standard exciton notation, $F^{L/U}$, where $F$ is the $z$-projection of 
the total exciton angular momentum and $L/U$ stands for lower/upper state with a given $F$.
In the spherical core limit, $R_z/R_\perp=1$, there are two subsets of exciton levels. 
The lower subset is formed by $\pm 2$ and $\pm 1$ doublets. $\pm 2$ are optically inactive states,
while $\pm 1^L$ emit planar polarized light orthogonal to the $c$-axis. 
All these states are essentially HH in nature. The upper subset of exciton levels is formed by LH states 
instead.  $0^L$ is optically inactive, $\pm 1^U$ emits planar polarized light and $0^U$ emits exclusively linearly 
($z$-)polarized light, because spin selection rules only allow the $C$-band component of the LH to recombine with the 
electron.\cite{EfrosPRB2} 
The energy splitting between HH and LH subsets is roughly given by $\Delta$.
Thus, HH and LH subsets cross at $R_z/R_\perp \approx 1.2$, similar to \ref{fig3}(a).
Qualitatively similar fine structure spectra are obtained for larger cores, but with smaller intra-subset splittings
due to the weaker Coulomb interaction (not shown).

 Notice in \ref{fig4}(a) that $0^U$ is never the ground state. Even for $R_z/R_\perp>1.2$, when $\Delta<0$,  $\pm 1^L$ states are lower in energy, and they emit planar polarized light. Thus, a net linearly polarized exciton emission requires thermal energy. Negative $\Delta$ values help increase the thermal occupation of $0^U$, but a for a complete understanding of room temperature polarization one has to compare and analyze the emission from several competing states.
 We then calculate the polarization degree:
\begin{equation}
p= \frac{I_z - I_\perp}{I_z + I_\perp},
\label{pol}
\end{equation}
\noindent where $I_z$ is the intensity of light emitted along the $c$-axis, given by: 
\begin{equation}
I_z = N_{0^U}\, R^e_z \, f_{0^U}
\end{equation}
\noindent and $I_\perp$ that in an arbitrary direction perpendicular to it:
\begin{equation}
I_\perp = 2\times\,R^e_\perp \left(N_{\pm 1^L}\, f_{\pm 1^L} + 
                        N_{\pm 1^U}\, \, f_{\pm 1^U} \right). 
\label{eq:Iperp}
\end{equation}
\noindent In the above expressions, $N_{F^{U/L}}$ is the room temperature occupation of the 
state $F^{U/L}$, $f_{F^{U/L}}$ the corresponding oscillator strength, and $R^e_z$ ($R^e_\perp$) 
the local field parameter, accounting for the dielectric screening of electromagnetic fields along the $c$-axis 
(orthogonal to it). Since the last term depends on the aspect ratio of the shell only, we choose 
DiRs with fixed aspect ratio of 10, so that all differences we may observe arise solely from 
electronic factors. The factor 2 in Eq.~(\ref{eq:Iperp}) accounts for the state degeneracy.

\ref{fig4}(b) shows the calculated degree of polarization for DiRs with different geometries.
Important differences are observed, which shows that $p$ is not governed by the shell aspect ratio 
alone, as sometimes suggested.\cite{CoropceanuACS} 
 Given a fixed core size, linear polarization is much more pronounced for thin CdS shell (solid lines) 
than for thick shell (dashed lines), in agreement with the experiments of Refs.~\cite{DirollJPCL,VezzoliACS}. 
Besides, core ellipticity is found to have a major influence. 
The highest degrees of linear polarization are obtained for prolate core no matter the shell thickness. 
This is consistent with experiments showing that rod-in-rods have $p$ values systematically larger than DiRs.\cite{SittNL,HadarJPCL}
 Both these effects can be understood from the geometry-induced decrease of $\Delta$, which translates into larger $N_{0^U}$.
Note however that strong polarization $p$ can be achieved even if $\Delta > 0$. For example, a spherical core with 
$(R_\perp,H)=(1.0,0.5)$ nm has $p=0.71$ in spite of having $\Delta=30$ meV. This shows that even for relatively
low $N_{0^U}$, dielectric effects and the relative oscillator strengths $f_{F^{U/L}}$ can lead to net linear polarization.
 
 On the other hand, \ref{fig4}(b) also shows that larger cores generally show higher $p$ values than small ones.
This is in close agreement with the the central experimental observation of Ref.~\cite{VezzoliACS},
namely that the core size is a chief factor determining $p$ in DiRs, but again it is beyond purely 
energetic interpretations, since --as shown on \ref{fig3}(a)-- large cores require higher core aspect ratio 
for LHs to become ground state. 
%
%
 The interpretation of this phenomenon is as follows. In large cores,
the oscillator strength $f_{0^U}$ is much larger than that of other thermally populated
states emitting $(x,y)$-polarized light. This is illustrated in Fig.~\ref{fig4}(c) for DiRs with $(R_\perp,H)=(2,0.5)$ nm: 
if the core is spherical, $f_{0^U}$ is 4-6 times larger than $f_{\pm 1^L}$ and $f_{\pm 1^U}$, 
and the difference increases as the core becomes prolate.
 The underlying reason can be seen in \ref{fig3}(e):  shear strain-induced PZ in large core DiRs leads to stronger 
(weaker) electron-LH (HH) overlap, thus stimulating emission from $0^U$ (LH) relative to $\pm 1^U$ (HH).  
 In general, this effect scales with core size and ellipticity (see Figs.~S5 and S6 in the Supporting Information).
The faster recombination of LH excitons implies that pronounced linear polarization
can be achieved even if the ground state is HH.

To further evidence the critical role of strain DP and PZ in determining the optical anisotropy of DiRs, 
in \ref{fig4}(d-f) we show the fine structure spectrum, degree of linear polarization and relative oscillator strengths
obtained neglecting core/shell strain. One can see that the ground state remains HH,
 $f_{\pm 1^L}$ is closer to $f_{0^U}$ and, consequently, the polarization value $p$ is 
much smaller than in the strained case.  In fact, small radii now imply stronger linear polarization, 
in sharp contradiction with Ref.~\cite{VezzoliACS} experiments. 
It follows strain is crucial.\cite{OS_Vezzoli}


In summary, we have shown theoretically that linearly polarized light emission and absorption in core/shell CdSe/CdS DiRs
has an electronic origin associated with (i) the energetic stabilization of LHs with respect to HHs, and (ii) the
enhanced recombination rate of LH excitons compared to HH ones. 
 The energetic stabilization of LHs takes place upon the growth of the CdS shell around the CdSe seed,
mainly via shear strain, with additional contributions from electron-hole Coulomb interaction 
and quantum confinement.  
Shear strain, $\varepsilon_\perp - \varepsilon_{zz}$, is roughly zero inside spherical CdSe cores,
 but  wurtzite cores tend to grow slightly prolate along the $c$-axis. 
In such a case, $\varepsilon_{zz}$ becomes more compressive  and  $(\varepsilon_\perp - \varepsilon_{zz})>0$,
 which stabilizes LHs relative to HHs via deformation potential. 
A thin diameter CdS shell has a similar --albeit weaker-- influence on shear strain, 
as it enables the core to dilate in the transversal direction and hence relax $\varepsilon_\perp$.
 The enhanced recombination rate of LH excitons is in turn related to shear strain-induced PZ,
which becomes important in DiRs with large, prolate core.\cite{SegarraJPCL}
Piezoelectric fields push electron and hole wave functions towards opposite sides of the core.
 This effect is more pronounced for HHs than for LHs, owing to their heavier masses along the $c$-axis.
 Consequently, $0^U$ states have faster radiative recombination rates 
than $\pm 1^L$ and $\pm 1^U$, and room temperature emission becomes linearly polarized.

From our analysis, it follows that the geometrical conditions optimizing linear polarization in DiRs 
are prolate core shape (enhanced shear strain, large PZ dipole and electron-LH attraction), 
large core size (large PZ dipole, weak confinement), and thin shell (enhanced shear strain).
 These findings support  and provide a unified explanation for recent experimental works, 
which phenomenologically inferred similar geometric conditions.\cite{VezzoliACS,SittNL,HadarJPCL,DirollJPCL}
 Together with the well-known dielectric origin of linear polarization in rod-shaped structures,\cite{ShabaevNL} 
our results complete the theoretical knowledge required for deterministic engineering of absorption and emission
anisotropy in DiRs.



\acknowledgement

Support from MINECO project CTQ2014-60178-P, UJI project P1-1B2014-24 is acknowledged.\\

{\bf Supporting Information Available:} Methods, material parameters, additional calculations on: the effect of the rod length; robustness against the use of different sets of masses and deformation potential parameters; CdSe seeds with no shell; dependence of e-LH/HH overlap vs core size in DiRs.


\begin{thebibliography}{32}
\bibitem{OS_Vezzoli}
A comparison between \ref{fig4}(c) and (f) shows that unstrained oscillator strengths
like those of Ref.~\cite{EfrosPRB2} are unsuited for analysis of DiRs. Thus, the
quantitive data fitting of Ref.~\cite{VezzoliACS} should be taken with care.
\end{thebibliography}
\end{document}




\newpage

\section{Methods}

Strain maps are calculated in the continuous medium model by minimizing the elastic energy. 
The boundary conditions are zero normal stress for the free surface.\cite{RajadellJAP} 
The strain tensor elements $\epsilon_{ij}(\mathbf{r})$, ensuing piezoelectricity polarization vector
and potential are obtained using the multiphysics mode of Comsol 4.2 software.
Electron states are calculated with a 3D single-band (actually two uncoupled bands including spin) 
effective mass Hamiltonian. Hole states are calculated with a 3D six-band Hamiltonian for wurtzite 
including $A$-band, $B$-band and $C$-band with spin-orbit interaction, as well as position dependent
mass (Burt-Foreman-like).\cite{ClimenteJAP}
The strain-induced deformation potential term is isomorphic to the kinetic energy term\cite{ChuangPRB}
and the piezoelectric potential is diagonal. Electron-hole exchange interaction is projected on the 
basis of wurtzite Bloch functions used for the kinetic energy Hamiltonian, see Eq.~(\ref{eq_blochWZ}).
Interacting electron and hole states are obtained by iterative resolution of the Schrodinger-Poisson equation,
 within a self-consistent Hartree procedure, taking into account the dielectric mismatch with the 
dielectric surroundings of the NC.  
 For the calculation of the dielectric screening factors ($R^e_{z}$ and $R^e_\perp$ in Eqs.(4) and (5) of the 
main text), we approximate the shell shape as an ellipsoid. This simplification allows us obtain analytical 
expressions\cite{ShivolaJN} while preserving our qualitative conclusions, since in Fig.~4 we compare DiRs with 
identical shell aspect ratio and all differences are connected to electronic effects only.

\section{Disentanglement of physical factors contributing to HH-LH splitting}

Our Hamiltonian for excitonic holes, prior to including electron-hole exchange interaction 
(see next section), reads: 
\begin{equation}
\label{eq1}
H = H_k + V_{qc} + V_{psp} + H_{dp} + V_{pz} + V_{e-h},
\end{equation}
\noindent where $H_k$ is the six-band k$\cdot$p wurtzite (WZ) Hamiltonian accounting for kinetic energy,\cite{ChuangPRB} 
$V_{qc}$ the (diagonal) confinement potential defined by heterostructure band offsets, $V_{psp}$ the (diagonal)
spontaneous polarization potential, $H_{dp}$ the six-band Hamiltonian accounting for strain-induced
deformation potential,\cite{ChuangPRB} $V_{pz}$ the (diagonal) piezoelectric potential and 
$V_{e-h}$ the (diagonal) Coulomb attraction exerted by the electron on the hole.

To disentangle the contribution of individual mechanisms to the energy splitting 
between heavy holes (HHs) and light holes (LHs), we define partial Hamiltonians:
\begin{eqnarray}
H_{shape}&=& H_k + V_{qc} + V_{psp},\\
H_{strain}&=& H_{shape} + H_{dp} + V_{pz},\\
H_{coulomb}&=& H_{strain} +  V_{e-h}.
\end{eqnarray}
\noindent $H_{shape}$ accounts for quantum confiement and spontaneous polarization effects only,
which are essentially defined by the core size and shape.
$H_{strain}$ adds strain via deformation potential and piezoelectricity.
Last, $H_{coulomb}$ adds interaction with the electron.
We run independent calculations with each Hamiltonian. In each case, the resulting 
HH-LH energy splitting can be identified with individual mechanisms as summarized in the table below:\\
\begin{table}[h]
\begin{tabular}{|c|c|}
\hline
{\bf Hamiltonian} & {\bf HH-LH splitting} \\
\hline
$H_{shape}$ & $\Delta_{shape} + \Delta_{int}$ \\ 
\hline
$H_{strain}$ & $\Delta_{strain} + \Delta_{shape} + \Delta_{int}$ \\
\hline
$H_{coulomb}$ & $\Delta_{coulomb} + \Delta_{strain} + \Delta_{shape} + \Delta_{int}$ \\
\hline
\end{tabular}
\end{table}

In the table, $\Delta_{int}$ is the intrinsic HH-LH splitting due to 
crystal field and spin-orbit interaction, which is already present in bulk materials 
($\Delta_{int}=23.4$ meV for CdSe).\cite{ChuangPRB}
The values of $\Delta_{shape}$, $\Delta_{strain}$ and $\Delta_{coulomb}$ we show in the paper
are obtained from the differences between HH-LH splittings in different calculations.


\section{Electron-hole exchange interaction in WZ}

Following Efros et al.\cite{EfrosPRB2}, the electron-hole exchange Hamiltonian is taken as:

\begin{equation}
\label{eq7}   
\hat H_{ex} = a_{ex} \delta({\mathbf r}_e-{\mathbf r}_h) \hat \sigma \cdot \hat J,
\end{equation}

\noindent where the parameter $a_{ex}=(2/3)\,\epsilon_{ex}\,(a_0)^3$, with $a_0$ being the lattice constant and $\epsilon_{ex}$ the exchange interaction strength. Because holes are confined inside the CdSe core, we take CdSe values: $a_{0}=4.3$  \AA~ and $\epsilon_{ex}=450$ meV.\cite{EfrosPRB2}
$\hat \sigma$ are the Pauli matrices, 
\begin{equation}
\label{eq4}     
\sigma_x=\left[ \begin{matrix} 0 & 1  \cr 1 & 0\end{matrix} \right] \;\;\;\;\; 
\sigma_y=\left[ \begin{matrix} 0 & -i  \cr i & 0\end{matrix} \right] \;\;\;\;\;
\sigma_z=\left[ \begin{matrix} 1 & 0  \cr 0 & -1\end{matrix} \right] \;\;\;\;\;  
\end{equation}
\noindent which account for the electron spin $s=1/2$. For valence band holes in ZB (ZB),  
one can use Bloch functions $|J,J_z\rangle$ with $T_d$ symmetry:
\begin{align}
\label{eq_blochZB}
|3/2,3/2\rangle &= -\frac{1}{\sqrt{2}}\,|(X+iY) \uparrow \rangle &
|3/2,-3/2\rangle &= \frac{1}{\sqrt{2}}\,|(X-iY) \downarrow \rangle \\
\nonumber
|3/2,1/2\rangle &= \sqrt{\frac{2}{3}} |Z \uparrow \rangle - \frac{1}{\sqrt{6}} | (X+iY) \downarrow \rangle &
|3/2,-1/2\rangle &= \sqrt{\frac{2}{3}} |Z \downarrow \rangle + \frac{1}{\sqrt{6}} | (X-iY) \uparrow \rangle \\
\nonumber
|1/2,1/2\rangle &= \frac{1}{\sqrt{3}} |Z \uparrow \rangle + \frac{1}{\sqrt{3}} | (X+iY) \downarrow \rangle &
|1/2,-1/2\rangle &= -\frac{1}{\sqrt{3}} |Z \downarrow \rangle + \frac{1}{\sqrt{3}} | (X-iY) \uparrow \rangle. 
\end{align}
If one restricts to HH and LH states, both have total angular momentum $J=3/2$. 
Then, one can use the corresponding matrix representations:
\begin{equation}
\label{eq9}     
J_x=\left[ \begin{matrix} 0 & \frac{\sqrt{3}}{2}&0&0  \cr \frac{\sqrt{3}}{2} & 0 &1 &0 \cr 0 & 1 & 0 &\frac{\sqrt{3}}{2} \cr 0&0&\frac{\sqrt{3}}{2}&0\end{matrix} \right] \;\;\;\;\; 
J_y=\left[ \begin{matrix} 0 & -i\,\frac{\sqrt{3}}{2}&0&0  \cr \frac{i\,\sqrt{3}}{2} & 0 &-i &0 \cr 0 & i & 0 &-i\,\frac{\sqrt{3}}{2} \cr 0&0&i\,\frac{\sqrt{3}}{2}&0\end{matrix} \right] \;\;\;\;\;
J_z=\left[ \begin{matrix} \frac{3}{2} &0&0 & 0  \cr 0&\frac{1}{2} &0&0 \cr 0&0&-\frac{1}{2}&0\cr 0&0&0&-\frac{3}{2}\end{matrix} \right] \;\;\;\;\;  
\end{equation}
\noindent and expand Eq.~(\ref{eq7}) by carrying out the Kronecker product $\sigma_i \otimes \mathbb J_i$, with $i= x,y,z$.
This would lead to a Hamiltonian equivalent to Eq.~(14) of Ref.~\cite{EfrosPRB2}, except for some phases
(we use Condon-Shortley convention).
In this work, however, we go beyond the cubic approximation and employ truly WZ Hamiltonians. 
To derive the corresponding exchange interaction
Hamiltonian, we consider that the Bloch functions of valence band holes we use are those of Chuang and Chang:\cite{ChuangPRB}
\begin{align}
\label{eq_blochWZ}
|A \uparrow\rangle & = -\frac{1}{\sqrt{2}} | (X+iY) \uparrow \rangle  &
|A \downarrow\rangle & =  \frac{1}{\sqrt{2}} | (X-iY) \downarrow \rangle  \\
\nonumber
|B \uparrow\rangle & =  \frac{1}{\sqrt{2}} | (X-iY) \uparrow \rangle  &
|B \downarrow\rangle & = -\frac{1}{\sqrt{2}} | (X+iY) \downarrow \rangle  \\
\nonumber
|C \uparrow\rangle & = |Z \uparrow \rangle &
|C \downarrow\rangle & = |Z \downarrow \rangle. 
\end{align}
\noindent The above functions are eigenfunctions of angular momentum $\hat L$ and spin $\hat s=1/2\hat \sigma$ operators.
Then, we can obtain the matrix ${\mathbb J}^{WZ}$ from $\hat J_i=\hat L_i+\frac{1}{2}\hat \sigma_i$, with $i=x,y,z$,
and expand the exchange Hamiltonian, Eq.~(\ref{eq7}), by  carrying out the Kronecker product with the Pauli matrices of electrons. 
The resulting matrix, spanned on the basis
\footnotesize
\begin{equation*}
{|\uparrow\rangle |A \uparrow\rangle, \,
|\uparrow\rangle |B \uparrow\rangle, \,
|\uparrow\rangle |C \uparrow\rangle, \,
|\uparrow\rangle |A \downarrow\rangle, \,
|\uparrow\rangle |B \downarrow\rangle, \,
|\uparrow\rangle |C \downarrow\rangle, \,
|\downarrow\rangle |A \uparrow\rangle, \,
|\downarrow\rangle |B \uparrow\rangle, \,
|\downarrow\rangle |C \uparrow\rangle, \,
|\downarrow\rangle |A \downarrow\rangle, \,
|\downarrow\rangle |B \downarrow\rangle, \,
|\downarrow\rangle |C \downarrow\rangle} 
\end{equation*}
\normalsize
\noindent is:
\begin{equation}
\hat H_{ex} = a_{ex} \delta({\mathbf r}_e-{\mathbf r}_h) \,
\left(
\begin{array}{cccccccccccc}
 \frac{3}{2} & 0 & 0 & 0 & 0 & 0 & 0 & 0 & 0 & 0 & 0 & 0 \\
 0 & -\frac{1}{2} & 0 & 0 & 0 & 0 & 0 & 0 & \sqrt{2} & 0 & 0 & 0 \\
 0 & 0 & \frac{1}{2} & 0 & 0 & 0 & \sqrt{2} & 0 & 0 & 0 & 0 & 0 \\
 0 & 0 & 0 & -\frac{3}{2} & 0 & 0 & 0 & 1 & 0 & 0 & 0 & \sqrt{2} \\
 0 & 0 & 0 & 0 & \frac{1}{2} & 0 & 1 & 0 & 0 & 0 & 0 & 0 \\
 0 & 0 & 0 & 0 & 0 & -\frac{1}{2} & 0 & 0 & 1 & 0 & \sqrt{2} & 0 \\
 0 & 0 & \sqrt{2} & 0 & 1 & 0 & -\frac{3}{2} & 0 & 0 & 0 & 0 & 0 \\
 0 & 0 & 0 & 1 & 0 & 0 & 0 & \frac{1}{2} & 0 & 0 & 0 & 0 \\
 0 & \sqrt{2} & 0 & 0 & 0 & 1 & 0 & 0 & -\frac{1}{2} & 0 & 0 & 0 \\
 0 & 0 & 0 & 0 & 0 & 0 & 0 & 0 & 0 & \frac{3}{2} & 0 & 0 \\
 0 & 0 & 0 & 0 & 0 & \sqrt{2} & 0 & 0 & 0 & 0 & -\frac{1}{2} & 0 \\
 0 & 0 & 0 & \sqrt{2} & 0 & 0 & 0 & 0 & 0 & 0 & 0 & \frac{1}{2}
\end{array}
\right).
\end{equation}

In our calculations, exciton states are obtained in a two-step process. In the first step, we obtain exciton states as electron-hole products including direct Coulomb interaction only. In the second step, we diagonalize the exchange Hamiltonian on the basis of the eight lowest exciton states. 

\section{Material parameters}

Below we summarize the material parameters used in the calculations.
$m_0$ is the free electron  mass and $\varepsilon_0$ the vacuum permitivitty. 
A relative dielectric constant of 3 and confining potential of 5 eV is taken outside the 
NC to account for the dielectric environment. 
See Ref.~\cite{ClimenteJAP} for the Burt-Foreman kinetic energy term of Hamiltonian (to avoid
spurious solutions, the hole mass parameters we use follow the complete asymmetric operator ordering, 
i.e. $A_i^{(+)}=A_i$ and $A_i^{(-)}=0$).

\begin{longtable}{|c|c|c|c|c|c|c|}
\hline
{\bf Description} &   {\bf Symbol}	& {\bf CdSe WZ}	& {\bf CdS WZ}	& {\bf Units}		& {\bf CdSe Ref.} & {\bf CdS Ref.}\\
\hline
Elastic modulus tensor  & $C_{11}$  	& $74.1 \cdot 10^9$ 	& $86.5 \cdot 10^9$	&  Pa	& \cite{Sadao_book} p.333 & \cite{Sadao_book} p.278 \\ 
Elastic modulus tensor  & $C_{12}$  	& $45.2 \cdot 10^9$ 	& $54.0 \cdot 10^9$	&  Pa	& \cite{Sadao_book} p.333 & \cite{Sadao_book} p.278\\ 
Elastic modulus tensor  & $C_{13}$  	& $38.9 \cdot 10^9$ 	& $47.3 \cdot 10^9$	&  Pa	& \cite{Sadao_book} p.333 & \cite{Sadao_book} p.278\\ 
Elastic modulus tensor  & $C_{33}$  	& $84.3 \cdot 10^9$ 	& $94.4 \cdot 10^9$	&  Pa	& \cite{Sadao_book} p.333 & \cite{Sadao_book} p.278\\ 
Elastic modulus tensor  & $C_{44}$  	& $13.4 \cdot 10^9$ 	& $15.0 \cdot 10^9$	&  Pa	& \cite{Sadao_book} p.333 & \cite{Sadao_book} p.278\\ 
\hline
Piezoelectric constant  & $e_{31}$	& $-0.16$		& $-0.24$		&  C$\cdot$m$^2$	&  \cite{Berlincourt} &  \cite{Berlincourt} \\
Piezoelectric constant  & $e_{33}$	& $0.347$		& $0.44$		&  C$\cdot$m$^2$	&  \cite{Berlincourt} &  \cite{Berlincourt} \\
Piezoelectric constant  & $e_{15}$	& $-0.138$		& $-0.21$		&  C$\cdot$m$^2$	&  \cite{Berlincourt}  & \cite{Berlincourt} \\
\hline
Spontaneous polarization & $P_s$	& $-0.006$		& $-0.002$		&  C$/$m$^2$	&  \cite{Schmidt}  & \cite{Jerphagnon} \\
\hline
Dielectric constant 	& $\varepsilon_{\perp}$ & $9.29$	& $8.28$		&  $\varepsilon_0$	& \cite{Geick} & \cite{Ninomiya} \\
Dielectric constant 	& $\varepsilon_{z}$ 	& $10.16$	& $8.73$		&  $\varepsilon_0$	& \cite{Geick} & \cite{Ninomiya} \\
\hline
Lattice constant $\parallel$ $c$ axis & $c$	& $7.01$		& $6.749$		&   \AA 		& \cite{Reever} & \cite{Woodbury} \\
Lattice constant $\perp$ $c$ axis & $a$	& $4.30$		& $4.135$		&   \AA 			& \cite{Reever} & \cite{Woodbury} \\
\hline
Conduction band offset 	& $cbo$		& $0.0$			& $0.200$		&  eV			& \cite{BrovelliNC} & \cite{BrovelliNC} \\
Valence band offset 	& $vbo$		& $0.0$			& $-0.409$		&  eV			& \cite{WeiJAP} & \cite{WeiJAP} \\ 
\hline
Crystal field splitting	& $\Delta_1$	& $0.039$		& $0.027$		&  eV			& \cite{SirenkoSSC} & \cite{SirenkoSSC}  \\
Spin-orbit matrix element & $\Delta_2$	& $0.139$		& $0.022$		&  eV			& \cite{SirenkoSSC}& \cite{SirenkoSSC}  \\
Spin-orbit matrix element & $\Delta_3$	& $0.139$		& $0.022$		&  eV			& \cite{SirenkoSSC} & \cite{SirenkoSSC} \\
\hline
Electron mass 		& $m_z^*$		    & $0.115$		& $0.198$		&  $m_0$		& \cite{Eaves}  & \cite{Huang} \\
Electron mass 		& $m_{\perp}^*$		& $0.12$		& $0.23$		&  $m_0$		& \cite{Eaves}  & \cite{Huang}\\
\hline
Hole mass parameter 	& $A_1$		& $-5.06$		& $-4.53$		& $1/m_0$		& \cite{SirenkoSSC} & \cite{SirenkoSSC}\\
Hole mass parameter 	& $A_2$		& $-0.43$		& $-0.39$		& $1/m_0$		& \cite{SirenkoSSC} & \cite{SirenkoSSC}\\
Hole mass parameter 	& $A_3$		& $4.5$			& $4.02$		& $1/m_0$		& \cite{SirenkoSSC} & \cite{SirenkoSSC}\\
Hole mass parameter 	& $A_4$		& $-1.29$		& $-1.92$		& $1/m_0$		& \cite{SirenkoSSC} & \cite{SirenkoSSC}\\
Hole mass parameter 	& $A_5$		& $-1.29$		& $-1.92$		& $1/m_0$		& \cite{SirenkoSSC} & \cite{SirenkoSSC}\\
Hole mass parameter 	& $A_6$		& $-0.47$		& $-2.59$		& $1/m_0$		& \cite{SirenkoSSC} & \cite{SirenkoSSC}\\
\hline
CB Deformation pot. $\parallel$ $c$ axis	& $a_c^{z}$	& $-1.52$		& $-5.6$		&  eV	& \cite{Langer}	& \cite{Langer} \\
CB Deformation pot. $\perp$ $c$ axis	& $a_c^{\perp}$	& $-0.46$ 		& $-6.0$ 		&  eV	& \cite{Langer}	& \cite{Langer} \\
VB Deformation pot. 	& $D_1$		& $-0.76$		& $-2.8$		& eV				& \cite{Langer}	& \cite{Langer} \\
VB Deformation pot. 	& $D_2$		& $3.24$ 		& $-1.5$ 		& eV				& \cite{Langer}	& \cite{Langer} \\
VB Deformation pot. 	& $D_3$		& $4.0$			& $1.3$		& eV				& \cite{Langer}	& \cite{Langer} \\
VB Deformation pot. 	& $D_4$		& $-2.2$		& $-2.9$			& eV				& \cite{Langer}	& \cite{Langer} \\
VB Deformation pot. 	& $D_5$		& $1.2$			& $-1.5$			& eV				& \cite{Langer}	& \cite{Langer} \\
VB Deformation pot. 	& $D_6$		& $1.5$			& $1.2$		& eV				& \cite{Langer}	& \cite{Langer} \\
\hline
\caption{\small Wurtzite CdSe and CdS parameters used in the calculations.}\label{t:params}
\end{longtable}

For the conduction band offset, different experiments on CdSe/CdS DiRs report different values 
from -0.25 eV to 0.30 eV (see Ref.~\cite{SteinerNL} and references therein). 
The most recent studies seem to confirm that band alignment is either type-I or quasi-type-II.\cite{SteinerNL,BrovelliNC,EshetNL} 
We then choose a positive value of 0.2 eV, which grants such an alignment by enabling a moderate 
degree of electron wave function penetration into the shell.
 With the above parameters, for a CdS shell with aspect ratio 10 we obtain dielectric screening factors
$R^e_z=0.927$ and $R^e_\perp=0.288$.

On a different note, a few comments are worth regarding the choice of deformation potentials.
The deformation potentials $C_i$ in  Langer\cite{Langer} correspond to exciton, i.e., $C_i= a_i-D_i$. For sake of symmetry, only those conduction deformation potentials related to diagonal strain $\varepsilon_{zz}$ ($a_1=a_c^z$) and $\varepsilon_{\perp}$ ($a_2=a_c^{\perp}$) do not vanish. Care should be also taken with the phase factors. Since we employ the rather standard phases of Chuang and Chang WZ Hamiltonian,\cite{ChuangPRB} the comparison with that employed by Langer for exciton (see the exciton Hamiltonian in p. 4014 of Ref.~\cite{Langer}) leads to the following relationships:  $D_1=a_c^z-C_1$, $D_2=a_c{^\perp}-C_2$, $D_3=-C_3$, $D_4=-C_4$, $D_5=C_5$ and $D_6=C_6/2$. It is difficult, though, to experimentally disentangle conduction and valence deformation potentials. Then, one cannot find $a_c^z$ and $a_c^{\perp}$ for CdSe and CdS in the literature.
Actually, S.H. Park and Y.H. Cho provided such coefficients in Ref.~\cite{ParkJAP}. The authors quote M. Tchounkeu et al.\cite{TchounkeuJAP} 
as the source of these deformation potentials, but the use of the source data is unclear, as Ref.~\cite{TchounkeuJAP} does not disentangle conduction and valence deformation potentials. 

 Given that we have no reliable source for CdSe and CdS conduction deformation potentials, we approximate them as follows. On the one hand, it appears reasonable to assume that compounds in the WZ structure will have similar pressure coefficients as in the ZB structure, since the nearest-neighbor tetrahedral environment is similar in both structures. So is expected for the volume deformation potential $a$, related to the pressure coefficient through the bulk modulus B. As a matter of fact, Wei and Zunger\cite{wei_zunger} compare the LDA calculated pressure coefficients for AlN, GaN, and InN in the ZB and the WZ structures finding out negligible differences and then, employ the cubic model for all hexagonal studied compounds.
On the other hand, the distribution of the hydrostatic pressure shift between the conduction and valence bands $a = a_c - a_v$ found by them in the case of  CdS and CdSe, under the assumption of a cubic model, yields $a_c$ about twice than $a_v$  (the exact ratios are 2.1 and 2.4 for CdSe and CdS, respectively). In hexagonal symmetry (WZ) the isotropic character does not hold and so we should replace $a_c \; tr\,\epsilon$ by $a_c^z  \epsilon_{zz} +a_c^{\perp} \epsilon_{\perp}$. Since, as pointed out above, no conduction-valence distributed of $a_c^{(z,\perp)}$ data are available, we will assume the same conduction-valence distribution for $a_c^z$ and $a_c^{\perp}$ and, additionally, that it would be similar to that in cubic symmetry (i.e., conduction twice than valence). Then, we derive from this assumption the deformations potentials for conduction and valence band of CdS and CdSe by setting: $a_c^z = 2 D_1$ i $a_c^{\perp} = 2 D_2$. Then, $C_1=a_c^z-D_1=2 D_1-D_1=D_1$. In a similar way, $C_2=D_2$, etc. 

\section{Cubic deformation potential}
\label{s:cubic}

Although our calculations are carried out using WZ Hamiltonians,\cite{ChuangPRB} 
for clarity of the discussion in the paper we analyze the energetic splitting between HH and LH in terms of 
quasi-cubic deformation potentials terms. In this approximation, holes are modeled with the Hamiltonian of ZB
grown along [111]. When projected on the HH and LH states of Eq.~(\ref{eq_blochZB}), $|3/2,J_z\rangle$ 
with $J_z=3/2,1/2,-1/2,-3/2$, the kinetic term reads:
\begin{equation}
H_{LK} = -\frac{\hbar^2}{2\,m_0}\,
\left(
\begin{array}{cccc}
P+Q & -S & R & 0 \\
-S^\dagger & P-Q & 0 & R \\
R^\dagger & 0 & P-Q & S \\
0 & R^\dagger & S^\dagger & P+Q .
\end{array}
\right).
\end{equation}
\noindent where:
\begin{eqnarray}
P&=&\gamma_1 \,(k_x^2+k_y^2+k_z^2),\\
Q&=&\gamma_3 \,(k_x^2+k_y^2-2k_z^2),\\
R&=&-\frac{1}{\sqrt{3}}\,(\gamma_2+2\gamma_3)\,k_-^2 + \frac{2\sqrt{2}}{\sqrt{3}}\,(\gamma_2-\gamma_3)\,k_+\,k_z,\\
S&=&-\frac{\sqrt{2}}{\sqrt{3}}\,(\gamma_2-\gamma_3)\,k_+^2 + \frac{2}{\sqrt{3}}\,(2\gamma_2+\gamma_3)\,k_-\,k_z.
\end{eqnarray}
\noindent where $\gamma_i$ ($i=1,2,3$) are the Luttinger paramters.
Notice that the splitting between HHs ($|3/2,3/2\rangle$) and LHs ($|3/2,1/2\rangle$), neglecting band coupling, 
is given by $2Q$.

The strain deformation potential Hamiltonian is isomorphic to $H_{LK}$, and can be obtained with the following replacements: 
\begin{eqnarray}
\frac{\hbar^2}{2m_0}\,\gamma_1 & \rightarrow & -a_v, \\
\frac{\hbar^2}{2m_0}\,\gamma_2 & \rightarrow & -\frac{b}{2}, \\
\frac{\hbar^2}{2m_0}\,\gamma_3 & \rightarrow & -\frac{d}{2\sqrt{3}}.
\end{eqnarray}
\noindent where $a_v$, $b$ and $d$ are cubic deformation potentials. Then, the HH-LH energy splitting arising from
the strain deformation potential is $2q=\frac{d}{\sqrt{3}}\,(\varepsilon_{xx}+\varepsilon_{yy}-2\varepsilon_{zz})$,
with $\varepsilon_{ij}$ the strain tensor components.
There are no experimental values for $d$ in cubic CdSe, but we can estimate a value from the relationship
between cubic and hexagonal deformation potential parameters:\cite{ChuangPRB}
\begin{eqnarray}
D_1 &=& a_v + \frac{2d}{\sqrt{3}}, \\
D_2 &=& \frac{1}{3}\,(3 a_v - \sqrt{3} d),\\
D_3 &=& -\sqrt{3}\,d,\\
D_4 &=& \frac{\sqrt{3}}{2}\,d,\\
D_5 &=& \frac{b}{2}+\frac{d}{\sqrt{3}},\\
D_6 &=& \frac{6b + \sqrt{3}d}{3\sqrt{2}}.
\end{eqnarray}
Knowing the values of $D_1$ to $D_6$ (see \ref{t:params}), we calculate the values of $a_v$, $b$ and $d$ providing the best fit.
For CdSe (CdS), this gives us $d=-1.11$ eV ($d=-0.74$ eV), which is the value used in the paper. 
Please note that different values of $d$ are obtained using different partitions of the excitonic deformation potentials
$C_1$ and $C_2$ between conduction and valence band (see discussion in Material Parameter section).
Also, we find the cubic approximation in this case seems less accurate than in other materials (e.g. some nitrides).
Yet, for our qualitative analysis of results, the important feature is that $d$ is negative.
feature have checked this feature holds as well for other heuristic partitions of the coefficients, 
namely those proposed in \ref{figS3} below.

\section{Supporting calculations}

\subsection{Effect of rod length}

\begin{figure}[h]
\includegraphics[width=0.80\textwidth]{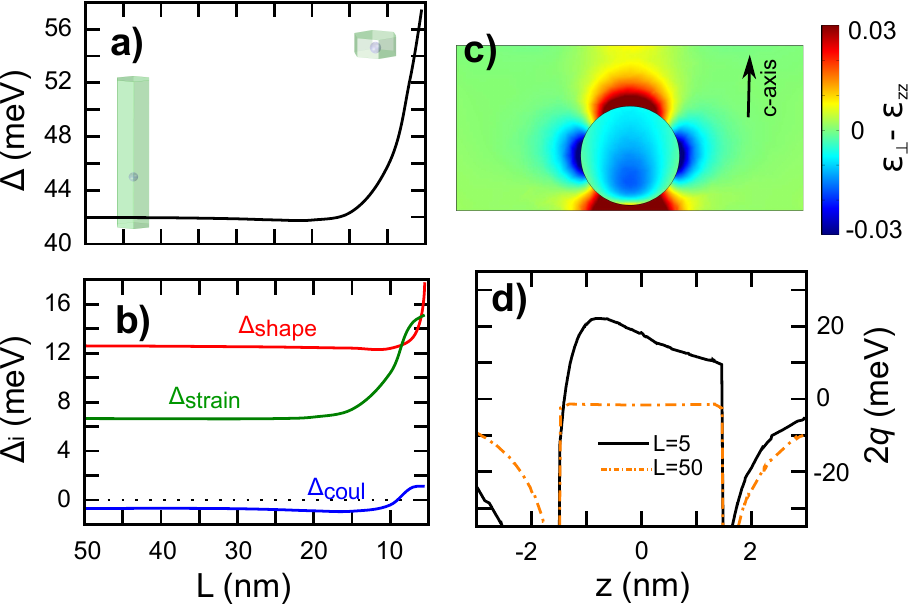}
\caption{(a-b) Same as Fig.2 of the paper but fixing the shell thickness ($H=3.5$ nm) and varying the length instead.
The HH-LH splitting is only affected when rods are very short (dot-in-plate limit). In such a case, HH becomes stabilized.
(c) shear strain in a dot-in-plate. 
(d) Offset between deformation potentials seen by HHs and LHs along the rod axis in the short shell case.
For $L=5$ nm HHs become stabilized.
In all calculations the core is spherical with $R=1.5$ nm.} 
\label{figS1}
\end{figure}

\ref{figS1}(a) shows the HH-LH splitting $\Delta$ in a DiR with fixed core radius and fixed shell thickness as a function of the rod length.
The splitting is only affected for small values of $L$ ($L<15$ nm). The effect of a short rod is the opposite to that of a thin rod
seen in Fig.2 of the paper. Namely, it makes $\Delta$ increase. One can see in \ref{figS1}(b) that this is mainly a strain effect.
The interpretation is again related to shear strain, $(\epsilon_{\perp} - \epsilon_{zz})$.
The strain deformation potential modifies $\Delta$ by an amount 
$2q=2\,d/\sqrt{3}\,\left( \epsilon_{\perp} - \epsilon_{zz} \right)$.
\ref{figS1}(c) shows that the shear strain in the core is negative, contrary to the case of thin but long shell studied in Fig.2(d).
The reason is that, for short shells, there are only a few monolayers of CdS on top and bottom of the core along the $c$-axis.
This allows the core to relax $\varepsilon_{zz}$, leading to $|\varepsilon_\perp| > |\varepsilon_{zz}|$.
Consequently, the deformation potential stabilizes HH relative to LH, \ref{figS1}(d). 
 
It is worth noting that the limit of small $L$ values we study in \ref{figS1} corresponds to the dot-in-plate structures
investigated in Ref.~\cite{CassetteACS}. Our findings are consistent with their experimental observation that HH-LH splitting
is enhanced in these objects as compared to dot-in-dots or DiRs. Likewise, our simulations confirm their interpretation
that the anisotropic strain induced by shell is responsible for the large $\Delta$ values. However, we estimate that varying
the shell anisotropy leads to changes in $\Delta$ of about 10 meV only. This is excellent agreement with the experiment, 
unlike the numerical results of their simple strain model, which estimated changes of more than 50 meV were 
theoretically possible. The large variations of $\Delta$ in their model are because they modeled the core as a cylinder
with infinitely thin shell on top and bottom. As compared to the realistic case of nearly-isotropic core with finite shell,
this exaggerates the values of shear strain. One should then expect weaker influence of the shell anisotropy
than suggested by their theory.

\subsection{Robustness against different mass parameters}

In the paper we have employed bulk WZ mass parameters ($A_{1-6}$), as defined in \ref{t:params}.
For confined CdSe and CdS nanocrystals, other sets of mass parameters have been proposed. Here we test the
robustness of the HH-LH ground state crossover using an alternative set of masses. Namely, we have
used quasi-cubic masses for CdSe ($\gamma_1=1.66$ and $\gamma_2=\gamma_3=0.41$)\cite{LaheldPRB} and 
CdS ($\gamma_1=2.33$ and $\gamma_2=\gamma_3=0.817$),\cite{FonoberovPRB} which can be related to WZ 
massic parameters as:
\begin{eqnarray}
A_1 & = & -\gamma_1 - 4\gamma_3,\\
A_2 & = & -\gamma_1 + 2\gamma_3,\\
A_3 & = & 6\gamma_3,\\
A_4 & = & -3\gamma_3,\\
A_5 & = & -\gamma_2-2\gamma_3,\\
A_6 & = & -\sqrt{2}\,(2\gamma_2+\gamma_3),\\
A_z & = & \gamma_2 - \gamma_3.
\end{eqnarray}

\begin{figure}[h]
\includegraphics[width=0.40\textwidth]{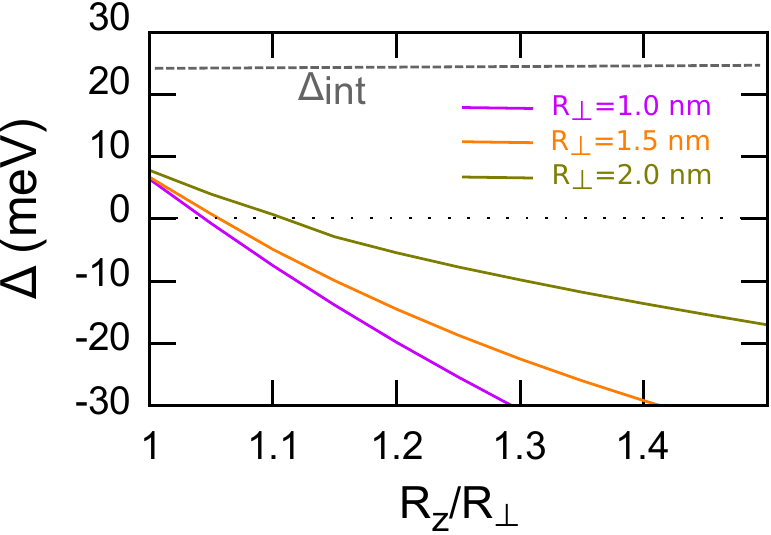}
\caption{Same as Fig.3(a) of the paper but using cubic masses for CdSe and CdS instead of WZ ones.
The same qualitative behavior is obtained.}
\label{figS2}
\end{figure}

\ref{figS2} shows the resulting HH-LH energy splitting as a function of the core aspect ratio.
One can see the trends are the same as with bulk WZ parameters (Fig.3(a) of the paper).
That is, for spherical cores the ground state is HH and, with increasing aspect ratio, it switches to LH ($\Delta < 0$).
Notice that the ground state crossover takes place for smaller aspect ratios than predicted with 
WZ masses ($R_z/R_\perp=1.05,\,1.06$ and $1.15$ for $R_\perp=1.0,\,1.5$ and $2.0$ respectively).
This is because of the different mass ratio between HHs and LHs.

\subsection{Robustness against different deformation potential parameters}

As mentioned in the Material Parameters section, the partition of band gap deformation potential $C_1$ and $C_2$
between conduction ($a^z_c,\,a^\perp_c$) and valence band ($D_1,\,D_2$) parameters is unknown in WZ CdSe and CdS. 
By analogy with Wei and Zunger\cite{wei_zunger}, in the paper we have assumed the conduction deformation 
potential is twice that of the valence band,  $a^z_c=2D_1$ and $a^z_\perp=2D_2$. Here we test different assumptions.

\begin{figure}[h]
\includegraphics[width=0.80\textwidth]{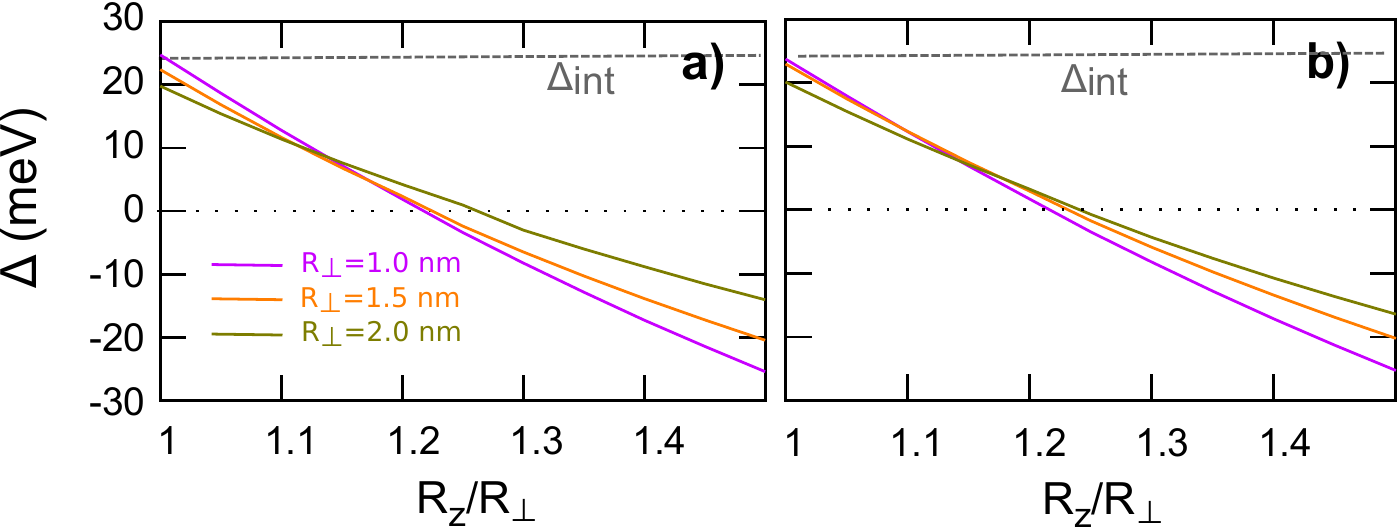}
\caption{Same as Fig.3(a) of the paper but using different partitions of deformation potentials $C_1$ and $C_2$.
(a) uses $a^z_c=-2D_1$ and $a^z_\perp=-2D_2$. 
(b) uses $a^z_c=2D_1$, $D_2=D_1+D_3$ and $a_c^{\perp}=a^{\perp}+D_2$ (see text).
The same qualitative behavior is obtained in all cases, with small numerical deviations.}
\label{figS3}
\end{figure}

In a first set of calculations, we consider  $a^z_c=-2D_1$ and $a^z_\perp=-2D_2$. This keeps conduction band parameters larger than valence
band ones (in absolute value), but accounts for the existing controversy on the sign of valence band deformation potentials.
In a second set of calculations, we take into account that in the case of nitrides, where the partition between conduction and valence
band deformation potentials is well known\cite{Vurgaftman}, the relationship $a_c^{z}/D_1 \approx 2$ holds reasonably well, 
but not for $a_c^{\perp}/D_2$. Then, more reasonable  $a_c^{\perp}$ and $D_2$ deformation potentials could be obtained 
from $D_1$ and $D_3$ by employing the cubic approximation, $D_2=D_1+D_3$, and then from $a_c^{\perp}=a^{\perp}+D_2$. 
The results with either set of parameters are plotted in \ref{figS3}. Again, they are qualitatively consistent with
those presented in the main paper.

\subsection{HH-LH splitting in CdSe quantum dots}

\begin{figure}[h]
\includegraphics[width=0.80\textwidth]{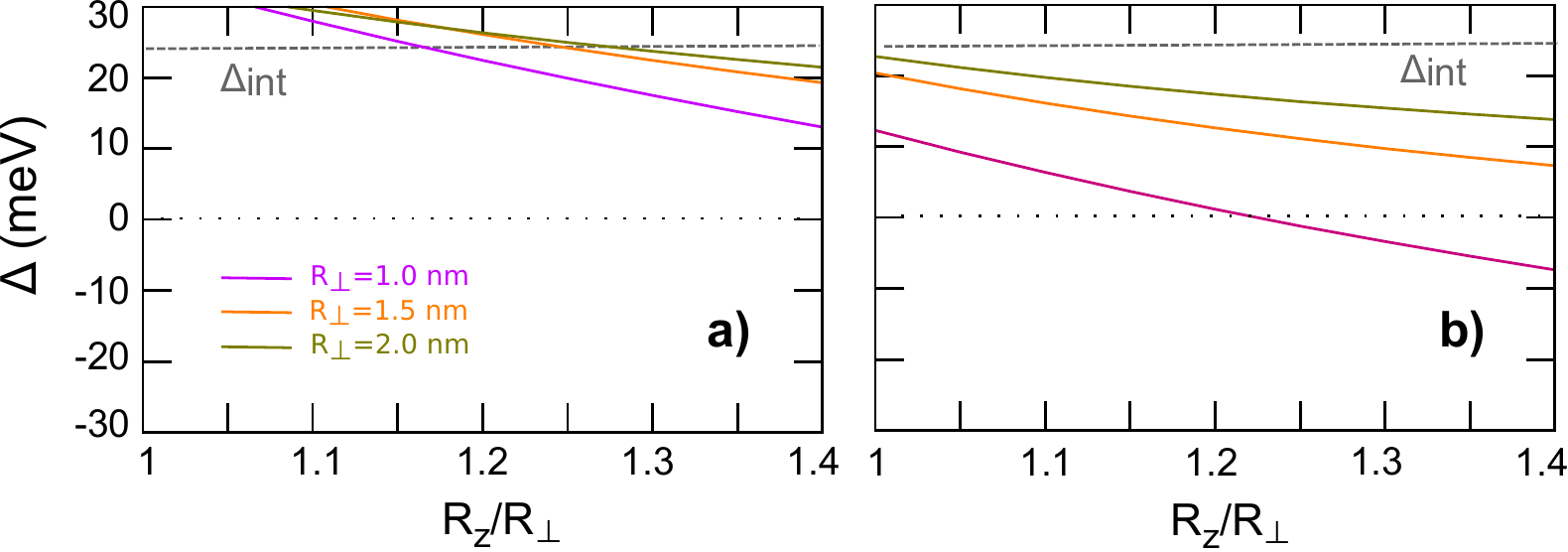}
\caption{(a) Same as Fig.3(a) of the paper but for CdSe quantum dot (core with no shell).
The critical aspect ratio for $\Delta$ to become negative is larger than that of CdSe seeds forming DiRs.
 (b) Same but using quasi cubic masses.} 
\label{figS4}
\end{figure}

 In Fig.~3(a) of the main text we show that LH ground states ($\Delta < 0$) are feasible in DiRs
with core aspect ratios between 1.1 and 1.3. This is largely due to shear strain induced by
the core/shell lattice mismatch. For comparison, in \ref{figS4}(a) we plot $\Delta$ for a CdSe core
with no shell. One can see $\Delta$ does not become negative for aspect ratios under 1.4. 
The comparison with DiRs evidences that the CdS shell plays a central role in the distinct behavior of DiRs, 
by greatly reducing the degree of core ellipticity needed to obtain LH ground states.
 
 It is worth mentioning that using quasi-cubic masses instead of bulk wurtzite ones, smaller critical
aspect ratios are obtained, as shown in \ref{figS4}(b). In fact, for $R_\perp=1$ nm these masses
predict a HH-LH crossover at $R_z/R_\perp = 1.22$, similar to estimates for CdSe nanorods obtained
with atomistic models.\cite{HuSCI} Nonetheless, the corresponding core aspect ratio in DiRs are 
again smaller than in the core only structure (cf. \ref{figS4}(b) with \ref{figS2}). Then, our
qualitative assessment on the role of strain is equally valid.

\subsection{Electron-hole overlap for HHs and LHs}

\begin{figure}[h]
\includegraphics[width=0.40\textwidth]{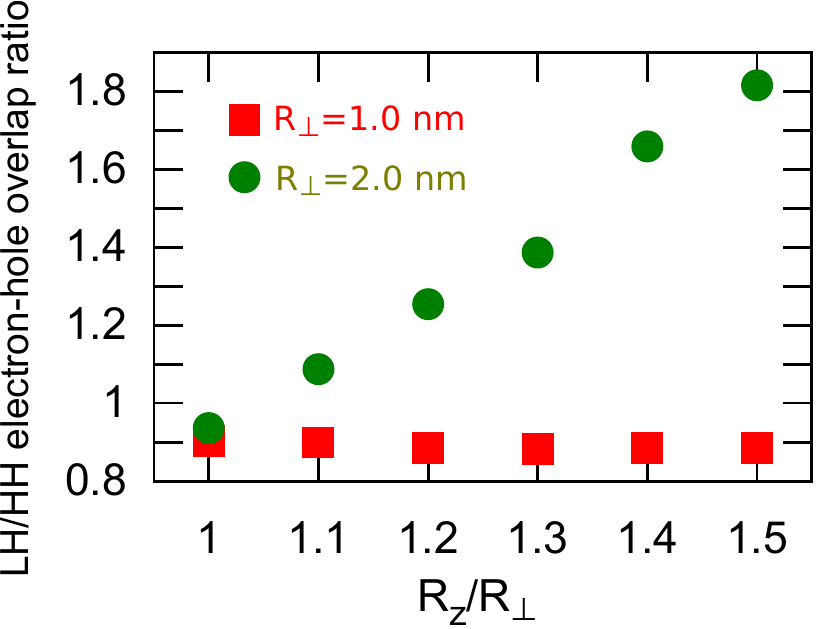}
\caption{Ratio of electron-hole overlap integrals for the lowest LH exciton vs lowest HH exciton
in DiRs with different core size and shape. Piezoelectricity leads to substantial differences in
large cores with prolate shape.}
\label{figS5}
\end{figure}

 In Fig.~3(e) of the main text we show that piezoelectricity pushes electron and hole
wave functions towards opposite sides of the core.  This impacts the radiative recombination
rate of HH excitons and LH excitons. As piezoelectricity becomes more important, the electron-hole
overlap (and hence the radiative recombination rate) decreases. However, because the wave function modulation
is more pronounced for HHs than for LHs (owing to their heavier mass), this also translates in
a relative increase of the LH exciton recombination rate against that of the HH exciton.
To illustrate this point, in \ref{figS5} we plot the ratio between LH and HH electron-hole overlap.
Piezoelectric dipoles increase with core size and ellipticity\cite{SegarraJPCL}. 
Consequently, the large core ($R_\perp=2$ nm), when prolate, displays electron-LH overlap almost
twice larger than of the HH. This has profound consequences on the linear polarization, as we 
discuss in the main text.

\subsection{Relative oscillator strenghts for small core DiRs}

\begin{figure}[h]
\includegraphics[width=0.50\textwidth]{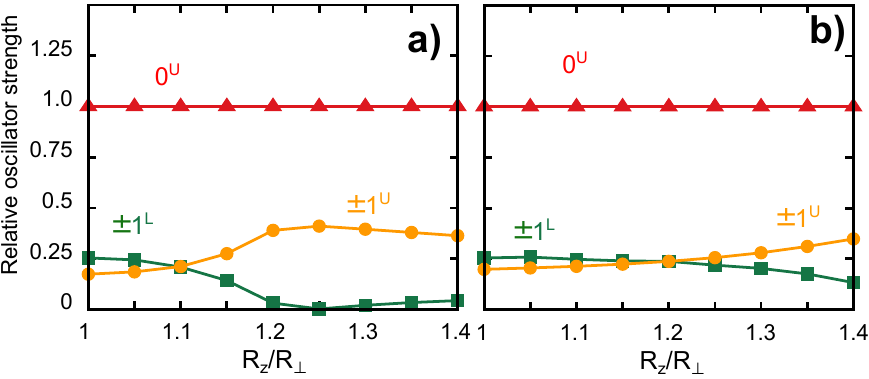}
\caption{(a) Oscillator strength of $\pm 1^{L/U}$ states with respect to that of $0^U$ in
DiRs with $R_\perp=1$ nm and $H=0.5$ nm. (b) Same but neglecting strain.}
\label{figS6}
\end{figure}

 \ref{figS6}(a) and (b) show the analogous of Fig.4(c) and (d) in the main text 
but for a DiRs with $R_\perp=1$ nm instead of $R_\perp=2$ nm.
Strain still leads to significant differences in the calculated oscillator strenghts.
For the strained case, \ref{figS6}(a), the oscillator strength of $\pm 1^{L/U}$ states is 
generally larger than that in Fig.~4(c). This is related to the weaker piezoelectricity in
small cores.

